\documentstyle[epsf]{aipproc}
\begin{document}
\title{Comparing electroweak data with a \\
 decoupling model}
\author{R. Casalbuoni$^{a,b,c}$, S. De Curtis$^b$, D. Dominici$^{a,b}$}
\author{R. Gatto$^c$ and M. Grazzini$^d$}
\address{$^a$Dipartimento di Fisica, Universit\`a di Firenze
\\ $^b$I.N.F.N., Sezione di Firenze
\\ $^c$D\'epartement de
Physique Th\'eorique, Universit\'e de Gen\`eve\\ $^d$Theoretical
Physics Division, CERN and  I.N.F.N., Gruppo collegato di Parma}

\maketitle

\def\MPL #1 #2 #3 {{\sl Mod.~Phys.~Lett.}~{\bf#1} (#3) #2}
\def\NPB #1 #2 #3 {{\sl Nucl.~Phys.}~{\bf B#1} (#3) #2}
\def\PLB #1 #2 #3 {{\sl Phys.~Lett.}~{\bf B#1} (#3) #2}
\def\PR #1 #2 #3 {{\sl Phys.~Rep.}~{\bf#1} (#3) #2}
\def\PRD #1 #2 #3 {{\sl Phys.~Rev.}~{\bf D#1} (#3) #2}
\def\PRL #1 #2 #3 {{\sl Phys.~Rev.~Lett.}~{\bf#1} (#3) #2}
\def\RMP #1 #2 #3 {{\sl Rev.~Mod.~Phys.}~{\bf#1} (#3) #2}
\def\ZPC #1 #2 #3 {{\sl Z.~Phys.}~{\bf C#1} (#3) #2}
\def\IJMP #1 #2 #3 {{\sl Int.~J.~Mod.~Phys.}~{\bf#1} (#3) #2}

\def\lsim{\mathrel{\raise.3ex\hbox{$<$\kern-.75em\lower1ex\hbox{$\sim$}}}}
\def\gsim{\mathrel{\raise.3ex\hbox{$>$\kern-.75em\lower1ex\hbox{$\sim$}}}}
\def\sigrts{\sigma_{\tiny\rts}^{}}
\def\sigrtssq{\sigma_{\tiny\rts}^2}
\def\sigrtsprime{\sigma_{E}}
\def\nsigrts{n_{\sigrts}}
\def\mupmum{\mu^+\mu^-}
\def\lplm{\ell^+\ell^-}
\def\drts{\Delta\sqrt s}
\def\rts{\sqrt s}
\def\ie{{\it i.e.}}
\def\eg{{\it e.g.}}
\def\eps{\epsilon}
\def\anti{\overline}
\def\wp{W^+}
\def\wm{W^-}
\def\mw{m_W}
\def\mz{m_Z}
\def\fbi{~{\rm fb}^{-1}}
\def\fb{~{\rm fb}}
\def\pbi{~{\rm pb}^{-1}}
\def\pb{~{\rm pb}}
\def\mev{~{\rm MeV}}
\def\gev{~{\rm GeV}}
\def\tev{~{\rm TeV}}
\def\spin{{2}}
\def\f{\frac}
\def\ct{c_{\theta}}
\def\st{s_{\theta}}

\def\pzero{P^0}
\def\mpzero{m_{\pzero}}
\def\pzerop{P^{0\,\prime}}
\def\gs{g^{\prime\prime}}
\newcommand{\be}{\begin{equation}}
\newcommand{\ee}{\end{equation}}
\newcommand{\bea}{\begin{eqnarray}}
\newcommand{\eea}{\end{eqnarray}}
\newcommand{\nn}{\nonumber}
\begin{abstract}
\noindent
Present data, both from direct Higgs search and from analysis of
electroweak data, are starting to become rather restrictive on
the possible values for the mass of the standard model Higgs. We
discuss a  new physics scenario based on a model with decoupling
(both in a linear and in a non linear version) showing how it
allows for an excellent  fit to the present values of the
$\epsilon$ parameters and how it widens the allowed ranges for
the Higgs mass (thought as elementary in the linear version, or
as composite in the non linear one).

\end{abstract}
\begin{center}

\vspace{4cm}
UGVA-DPT 1998/05-1006

University of Florence - DFF 310/05/98

\end{center}

\newpage

\section*{Introduction}
\noindent
The new LEP data presented at the recent Winter Conferences in
Moriond and La Thuile give  strong restrictions on the Higgs
mass. Direct Higgs search gives $m_H\ge 89.3~GeV$
 at 95\% CL \cite{L3}. From the global fit to all  electroweak
data one obtains $m_H\le 215~GeV$ at 95\% CL \cite{LEPWG}. The
corresponding bounds from the Jerusalem Conference of 1997
\cite{ward} were $77\le m_H(GeV)\le 420$ at 95\% CL. One reason
for the difference in the upper limit is the inclusion of the
most important part of the two-loop radiative corrections
\cite{sirlin}.

The upper bound on the Higgs mass comes mainly from the
experimental determination of  $\sin^2\bar\theta$ (where
$\bar\theta$ is the effective Weinberg angle). However there is
still a 2.3 $\sigma$ deviation between LEP and SLD averages. The
LEP average is by far dominated by the determination of
$A_{FB}^b$ which has the smallest experimental error. An upward
change of $\sin^2\bar\theta$ would increase the upper bound on
$m_H$, whereas a downward shift would lower it.

 The SLD average,
$\sin^2\bar\theta_{SLD}=0.23084\pm0.00035$, lies on the lower
side of the central value,
$\sin^2\bar\theta_{world~average}=0.23149\pm0.00021$, whereas the
LEP average, $\sin^2\bar\theta_{LEP}=0.23185\pm0.00026$, lies on
the higher side \cite{LEPWG}. Possible future experimental
results in the direction of lowering $\sin^2\bar\theta$ could
thus eventually lead to a conflict between the upper bound for
$m_H$ and the lower bound obtained from the direct search of the
Higgs. In such a  situation hints for physics beyond the standard
model would be obtained by looking at the $\epsilon$ parameters
\cite{altarelli1}. The ellipses in Figs. 1 and 2 are derived at $
1-\sigma$  from  all the latest electroweak data
\cite{caravaglios}.

 One notices that in these graphs  the
standard model points lie in general at higher values than  the
central experimental points, indicating a constraint
on the $\epsilon$ parameters to be smaller than the standard model
 values.

Not all models invoking new physics
would satisfy such a constraint. For instance,
elementary technicolor gives a contribution only to $\epsilon_3$,
but of the wrong sign.  The situation would be better for
supersymmetric models with  appropriate choices of the parameters
\cite{altarelli2}.

In this note we shall discuss the implications of a decoupling model
\cite{grazzini1,grazzini2} for new physics which presents the general
feature  of leading to contributions to all the $\epsilon$ parameters,
contributions all of negative sign.

By the requisite of decoupling, in a model for new physics, we
mean that the model is such that when the  mass scale for the
 new physics is made infinitely large the model goes back to the
standard model. The new mass scale controls the contributions to
the $\epsilon$ parameters.  The non linear
effective Lagrangian
model of ref. \cite{grazzini1} goes back for infinite mass scale
 to the standard model
without elementary Higgs. The renormalizable linear decoupling
model of ref. \cite{grazzini2} coincides for infinite mass
scale with the standard model, including its elementary Higgs,
at all perturbative orders.

\vskip1cm
\section*{The model}
\noindent
We will discuss the decoupling models described  in refs.
\cite{grazzini1,grazzini2}. The relation between the model
introduced in ref. \cite{grazzini1} and the one of
\cite{grazzini2} is analogous to the one between the non linear
and the linear $\sigma$-model. Both models  are based on the
gauge group $SU(2)_L\otimes U(1)\otimes SU(2)_{L}^\prime
\otimes SU(2)_{R}^\prime$ with gauge fields corresponding to
the ordinary gauge bosons $W^\pm$, $Z$ and $\gamma$ and new heavy
gauge fields ${\bf L}$ and ${\bf R}$. A discrete symmetry
$L\leftrightarrow R$ is also required such that the new gauge
fields have equal gauge couplings $g_L=g_R\equiv g_2$. The
symmetry also implies that at the lowest order in weak interactions
the masses of the new vector bosons are equal, $M_L=M_R\equiv M$.

The
gauge boson
masses are generated through the breaking of the gauge group down
to $U(1)_{\rm em}$, implying  9 Goldstone bosons.

In the non linear model \cite{grazzini1} these are all the scalar
fields. They all disappear  from the physical spectrum through
Higgs phenomenon.

In the linear version \cite{grazzini2} one introduces 3 complex
doublets belonging to the following representations of the global
group $SU(2)_L\otimes SU(2)_R\otimes SU(2)_{L}^\prime \otimes
SU(2)_{R}^\prime$
 \be
\tilde L\in (\spin,0,\spin,0),~~~~~
\tilde U\in (\spin,\spin,0,0),~~~~~
\tilde R\in (0,\spin,0,\spin)
\ee
These 3 doublets describe 9 Goldstone bosons and 3 physical neutral
scalar fields, one of which is the ordinary Higgs field
in the decoupling limit.

In the linear
model the breaking of the symmetry is supposed to come in two
steps characterized by the expectation values $\langle\tilde
L\rangle=
\langle\tilde R\rangle=u$ and $\langle\tilde U\rangle=v$ respectively.
The first two expectation values induce the breaking
$SU(2)_L\otimes SU(2)'_L\to SU(2)_{\rm weak}$ and $U(1)\otimes
SU(2)'_R\to U(1)_Y$, whereas the third one induces in the standard way
$SU(2)_{\rm weak}\otimes U(1)_Y\to U(1)_{\rm em}$. We assume that
the first breaking corresponds to a
scale $u\gg v$. In
the limit $u\to\infty$ the model decouples and one is left with
the  standard model with the usual Higgs \cite{grazzini1}.

One can think of the non linear version as the one to be used in
a scenario where the Higgs is thought  as composite  with a mass
at the $TeV$ scale. In ref. \cite{grazzini1} we have shown that
also the non linear model decouples.

A
feature of both models, the linear and the non linear one,
is that they  have an additional
accidental global symmetry $SU(2)\times SU(2)$, which acts
together with the usual $SU(2)$ to form a custodial symmetry. As
a consequence the new physics contribution to the $\epsilon$
parameters, at the lowest order in the weak interactions,
vanishes. In fact, the usual $SU(2)$  custodial requires the
vanishing of the contributions to $\eps_1$ and $\eps_2$, whereas
the new larger custodial symmetry implies also the vanishing of the
contribution to $\eps_3$.

Physically this is due to the mass and
coupling degeneracy between the new ${\bf L}$ and ${\bf R}$
resonances at the
lowest order. For this reason contributions to the $\epsilon$
parameters appear only to the next-to-leading
 order in the expansion in the
heavy masses. The tree-level contribution to the $\eps$
parameters at the first non trivial order in $1/M$
 is given for both linear and non linear version by
\cite{grazzini1,grazzini2}
\be
\Delta\eps_1 = -\f{\ct^4+\st^4}{\ct^2} X,~~~~~~~~
\Delta\eps_2 = - \ct^2 X,~~~~~~~~
\Delta\eps_3 = -X
\label{eps}
\ee
with $\theta$ the Weinberg angle. All the contributions are negative and are
all parametrized by the single parameter
\be
X=\left(\f g {g_2}\right)^2\f {M_Z^2}{M^2}
\label{X}
\ee
with $g$ the standard gauge coupling and $M_Z$ the $Z$ mass.

The linear model is renormalizable
and the corresponding radiative corrections can be evaluated by
following the lines of ref. \cite{grazzini2}.
The one-loop contribution to $\eps$ parameters
 is given by the usual radiative corrections of the standard
model plus the radiative corrections coming from new physics.
As
far as these last corrections are concerned,
 one can show (see \cite{grazzini2}) that, due to the
decoupling property,
they are typically smaller than  10\% of the tree-level
contributions. Therefore we will neglect them in our following
considerations since they are well below the experimental error
on the $\eps$ parameters which is of the order of $20\div30\%$.

The non linear model can be regularized assuming the linear model
as the regularizing theory and taking the Higgs mass as a cutoff
at the $TeV$ scale.

Therefore in both cases we get the same expressions for the
radiative corrections, except that in linear case the parameter
$m_H$ is  the physical Higgs mass, whereas in the non linear case
(where no elementary Higgs is present) one takes  $m_H$ as
describing a cutoff, to be chosen at around 1 $TeV$.

\vskip1cm
\section*{Comparison to electroweak data}
\noindent
As explained in the previous Section  the contributions  of new
physics to the $\eps$ parameters in the models considered here are all
negative and
parameterized in terms of the single variable $X$, which depends
on a combination of the  new mass scale $M$ and of the gauge
coupling of the new vector bosons $g_2$. In Figs. 1,2 we have drawn the
$1-\sigma$ experimental ellipses \cite{caravaglios} for the
 pairs
$(\eps_1,\eps_3)$ and $(\eps_3,\eps_2)$. The thick bars
correspond to the $\eps$ values of the standard model at given
Higgs mass ($m_H=70,~300,~1000~GeV$)  and with the top mass
varying in each case between $170.1$ and $181.1$ $GeV$ (from left to
right in Fig. 1 and from up to down in Fig. 2).

 For each
given pair of values of $m_t$ and $m_H$, one  considers a corresponding
line, parameterized by $X$ (see eq. (\ref{eps})), whose points give for
each $X$
 the values of the $\eps$  after inclusion of the new physics
discussed here. All these lines lie, in the figures, within the strips
attached to
each of the thick bars. For each  line originating from
the standard model points one can evaluate the best value for $X$
to fit the experimental values of the $\eps$ parameters. The
corresponding best fit points lie on the dashed bars of Figs. 1,2.

\begin{center}

\begin{tabular}{|c|c|c|c|c|c|c|c|c|c|}
\hline
 & \multicolumn{3}{|c|}{} & \multicolumn{3}{|c|} {}
& \multicolumn{3}{|c|}{} \\
$m_H$ &
\multicolumn{3}{|c|}{$\bar\epsilon_1\times 10^3$}&\multicolumn{3}{|c|}{$\bar
\epsilon_2\times 10^3$}
&\multicolumn{3}{|c|}{$\bar\epsilon_3\times 10^3$}
\\
(GeV)&
\multicolumn{3}{|c|} {$m_t$ (GeV) =} &  \multicolumn{3}{|c|}
{$m_t$ (GeV) =} & \multicolumn{3}{|c|} {$m_t$ (GeV) =} \\
& 170.1 & 175.6 & 181.1 & 170.1 & 175.6 & 181.1 & 170.1 & 175.6
 & 181.1\\\hline
 70 & 4.17 &  4.52    & 4.88 &  $-$8.44 & $-$8.77 &  $-$9.11 &
      3.62    &  3.42 &   3.22\\
100 & 3.94 &    4.29 &  4.65 &  $-$8.47 &   $-$8.79 &  $-$9.13 &
      3.83 &     3.63 &   3.43 \\
200 & 3.44 &    3.79 &  4.15 &  $-$8.41 &   $-$8.73 &  $-$9.06 &
      4.23 &     4.03 &   3.83 \\
300 & 3.13 &    3.47 &  3.83 &  $-$8.30 &   $-$8.61 &  $-$8.94 &
      4.47 &     4.27 &   4.07 \\
400 &  2.89 &   3.23 &  3.59 &  $-$8.19 &  $-$8.50 &  $-$8.83 &
      4.64 &     4.44 &   4.24 \\
500 &  2.67 &   3.04&   3.39 &   $-$8.09 & $-$8.39 &  $-$8.71 &
      4.77 &     4.57 &   4.37 \\
600 &  2.53 &   2.88 &  3.23 &  $-$7.99 & $-$8.29&   $-$8.61 &
      4.87 &     4.68 &   4.47 \\
700 &  2.39 &   2.74 &  3.09 &  $-$7.90 & $-$8.20 &  $-$8.51 &
      4.96 &     4.77 &   4.57 \\
800 &  2.27 &   2.61 &  2.96 &  $-$7.81 & $-$8.11  & $-$8.43 &
      5.04 &     4.85 &   4.64 \\
900 &  2.16 &   2.50 &  2.85 &  $-$7.73 & $-$8.03&   $-$8.34 &
      5.11 &     4.92 &   4.71 \\
1000 &  2.06 &  2.40 &  2.75 &   $-$7.66 &$-$7.95 &  $-$8.26 &
      5.18 &     4.98 &   4.78\\
\hline
\end{tabular}

\end{center}
\noindent
{\bf Table 1} - {\it
 $\eps$ parameters in the decoupling model
derived from the best fit value for $X$ for any pair $(m_t,m_H)$.
The experimental values for the
$(\epsilon_i)\times 10^3$ are:
 $\epsilon_1=3.85\pm1.20$,
$\epsilon_2=-8.3\pm 1.9$, $\epsilon_3=3.85\pm 1.21$} \cite{caravaglios}.
\vspace{1cm}

\noindent
The quality of the fit can be appreciated from Table 1 where we
give  the values of the $\eps$ parameters derived in each case
from the best value for $X$.
The best fit values for $X$ lie within $1.3\times 10^{-3}\div
2\times 10^{-3}$ for $170.1\le m_t(GeV)\le 181.1$ and
$70\le m_H(GeV)\le 1000$.

It is already clear from Figs. 1,2
 that, for any pair $(m_t,m_H)$, there is a value of $X$
which gives a fit to
the experimental data better than the one of the standard model.
This is emphasized in Fig. 3 where, for three
values of the top mass $m_t=170.1,~175.6,~181.1$ $GeV$, we plot,
as a function of the Higgs mass, the $\chi^2$ for the standard
model, and the $\chi^2$ for the decoupling model, where for each pair
$(m_t,m_H)$ the value of $X$ has been fixed at its best value.

This figure makes clear what we said before and shows also that
one gets an almost perfect agreement with the data in a region of
$m_H$ of order $100\div 300$ $GeV$.
This should be compared with
the result for the standard model which gives a best value for
$m_H$ of $66^{+74}_{-39}~GeV$ \cite{LEPWG}.
Correspondinlgy the 95\% CL bound on the Higgs mass goes from the
limit of 215 $GeV$, within the standard model, to values above 1 $TeV$ for
the models presented here.

\vskip1cm
\section*{Conclusion}
\noindent
The standard model fit to the electroweak data based on  the
latest experiments has considerably narrowed the allowed interval
for the standard Higgs particle mass. From the present situation
one might be afraid that the future increased experimental
accuracy could evidence a conflict between the lower bound on the
Higgs mass coming from the direct search of the particle and the
upper bound from the precision experiments. For this reason we
have discussed here the fit to the $\eps$ parameters in a
decoupling model (in two versions, a linear and a non linear one)
showing that within such a scenario such a  possible conflict
would be avoided. At the same time, an excellent fit is obtained
to the present determinations of $\eps$ parameters.

\vskip1cm
\noindent{\bf{Acknowledgement}}
\vskip0.5cm\noindent
We would like to thank G. Altarelli and F. Caravaglios for their
kindness in allowing us to use their new fit to the $\eps$
parameters.
This work is part of the EEC project "Tests of electroweak
symmetry breaking and future european colliders",
CHRXCT94/0579 (OFES 95.0200).

\begin{figure}[h]
\epsfxsize=15truecm.
\epsfysize=18truecm.
\centerline{\epsffile[40 200 480 680]{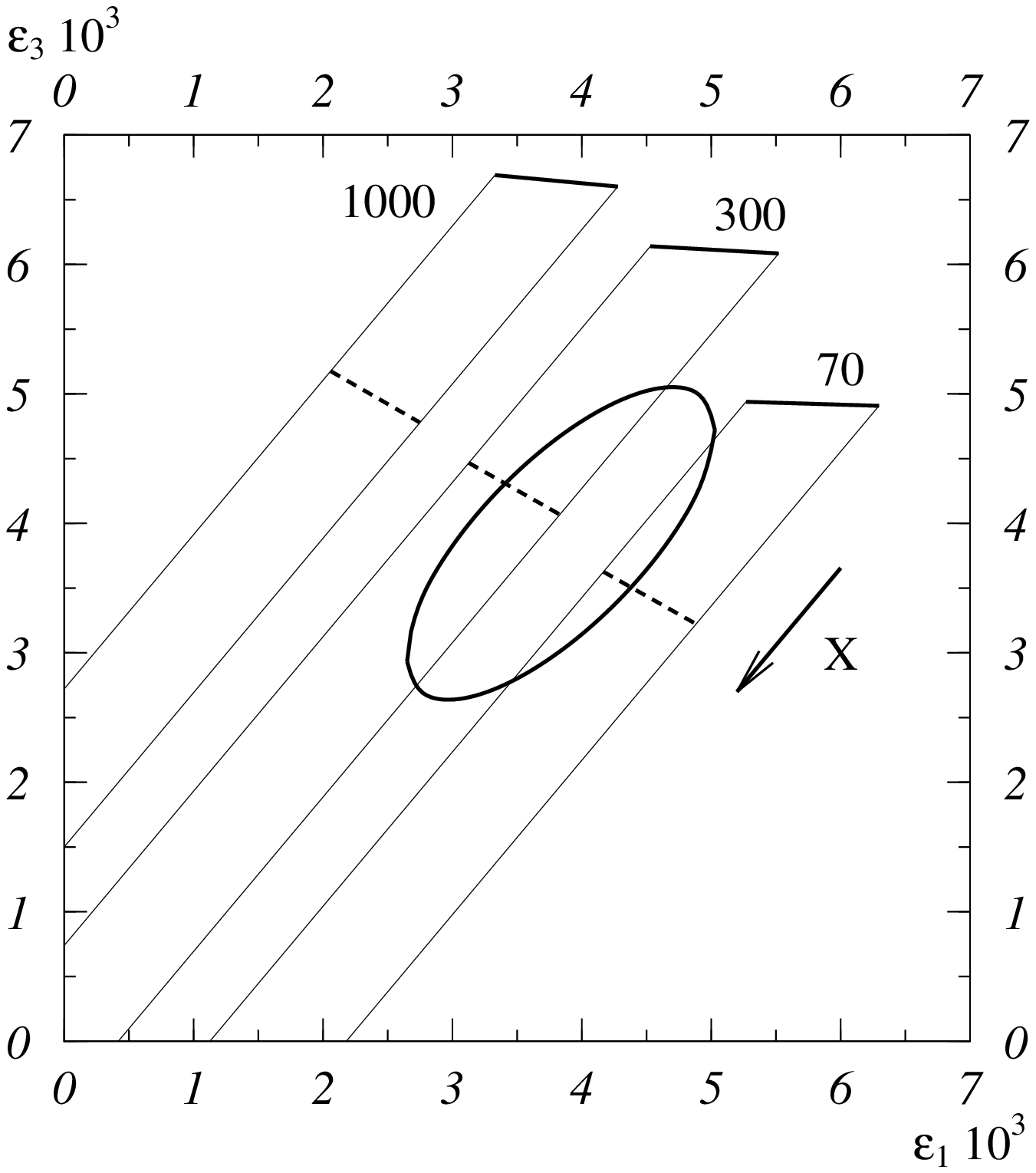}}
\noindent
{\bf {Fig. 1} }- {\it{Decoupling model predictions in the plane
$(\epsilon_1,\epsilon_3)$. The
ellipse corresponds to the $1-\sigma$ ($38\%$ probability
contour) experimental data. The thick continuous bars correspond
to the standard model predictions for $m_H=70,~300,~1000$ $GeV$,
and, for each case, $170.1\le m_t(GeV)\le 181.1$. For each choice
of the Higgs mass, the oblique strips correspond to  the
predictions of the model discussed here as parameterized by $X$,
see eq. (\ref{X}). The dashed bars describe, for each choice of
the Higgs mass, the best fits. }}
\end{figure}
\noindent

\begin{figure}[h]
\epsfxsize=15truecm.
\epsfysize=18truecm.
\centerline{\epsffile[40 200 480 680]{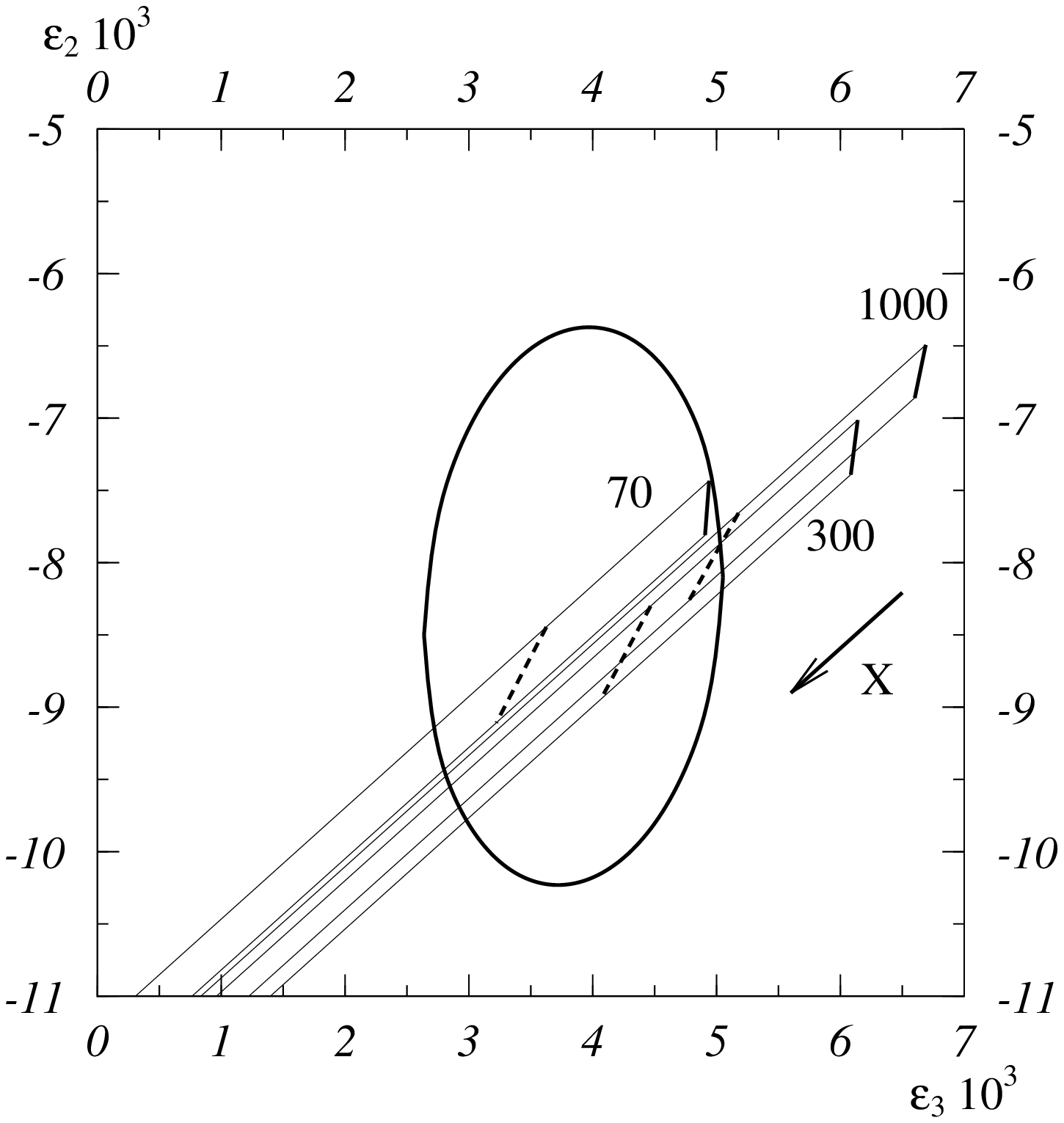}}
\noindent
{\bf {Fig. 2} }- {\it{
Decoupling model predictions in the plane
$(\epsilon_3,\epsilon_2)$. The graphical
representation is the same  as for Fig. 1.}}
\end{figure}
\noindent

\begin{figure}[h]
\epsfxsize=15truecm.
\epsfysize=18truecm.
\centerline{\epsffile[40 200 480 680]{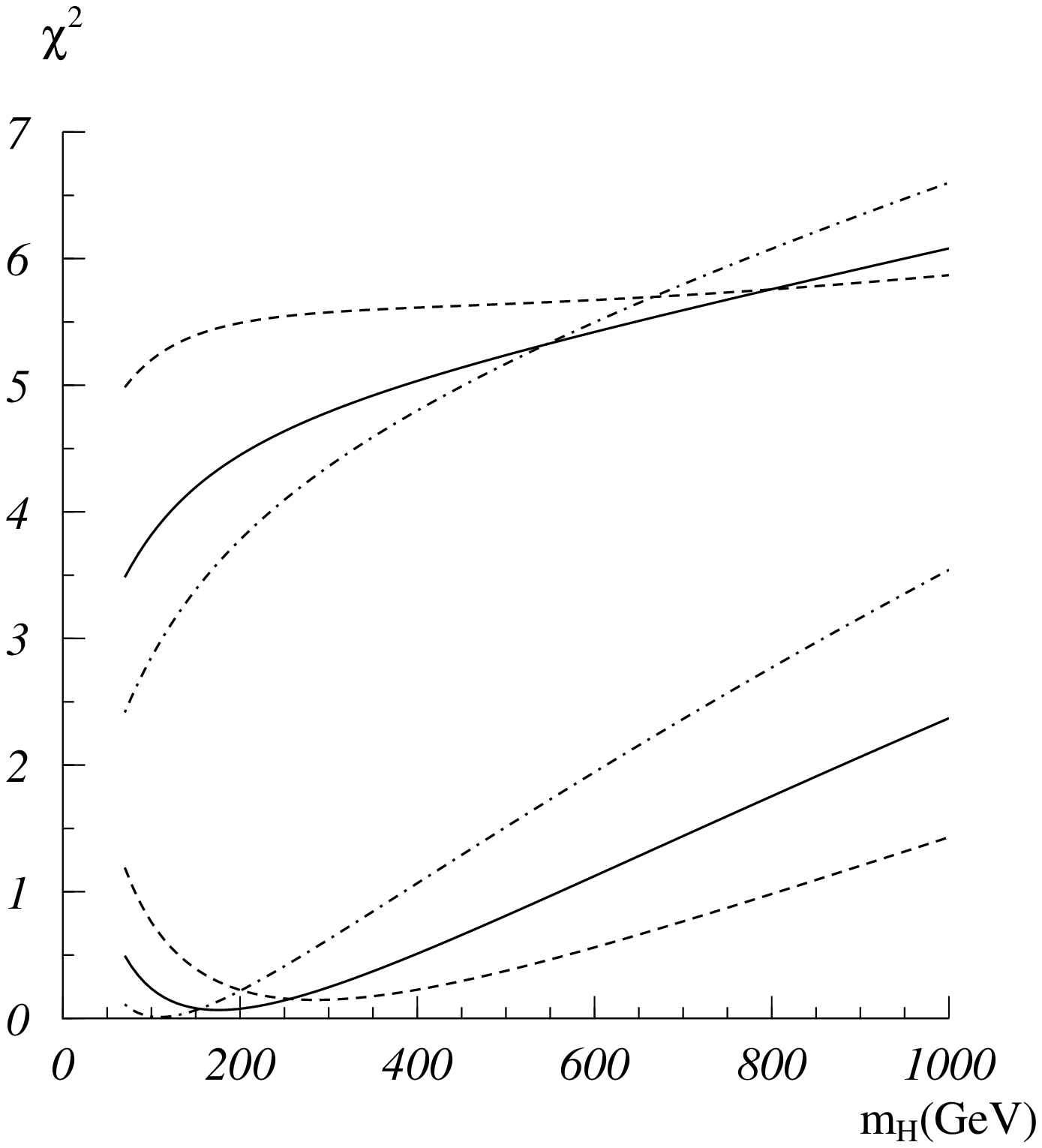}}
\noindent
{\bf {Fig. 3} }- {\it{$\chi^2$ vs. $m_H$. The curves  in the
upper part of the figure correspond to the standard model. Those
in the lower part correspond to the decoupling model  where at
each point the value of the parameter $X$ (see eq. (\ref{X})) is
that of the best fit. The continuous lines are for $m_t=
175.6~GeV$, the dash-dotted lines are for $m_t= 170.1~GeV$, and
the dashed ones are for $m_t= 181.1~GeV$.}}
\end{figure}

\end{document}